\documentclass[aps,reprint,twocolumn,superscriptaddress]{revtex4-2}
\usepackage{amsmath}
\usepackage{amssymb}
\usepackage{graphicx}
\usepackage{dcolumn}
\usepackage{bm}

\begin{document}

\preprint{APS/123-QED}

\title{Merging bound states in the continuum at third-order $\Gamma$ point enabled by controlling Fourier harmonic components in lattice parameters}%

\author{Sun-Goo Lee}
\email{sungooleee@gmail.com}
\affiliation{Department of Data Information and Physics, Kongju National University, Gongju, 32588, Republic of Korea}%
\affiliation{Institute of Application and Fusion for Light, Kongju National University, Cheonan, 31080, Republic of Korea}%

\author{Seong-Han Kim}
\affiliation{Advanced Photonics Research Institute, Gwangju Institute of Science and Technology, Gwangju 61005, Republic of Korea}%

\author{Wook-Jae Lee}
\email{wookjaelee@kongju.ac.kr}
\affiliation{Department of Data Information and Physics, Kongju National University, Gongju, 32588, Republic of Korea}%
\affiliation{Institute of Application and Fusion for Light, Kongju National University, Cheonan, 31080, Republic of Korea}%

\date{\today}

\begin{abstract}
Recent studies have demonstrated that ultrahigh-$Q$ resonances, which are robust to fabrication imperfections, can be realized by merging multiple bound states in the continuum (BICs) in momentum space. The merging of multiple BICs holds significant promise for practical applications, providing a robust means to attain ultrahigh-$Q$ resonances that greatly enhance light-matter interactions. In this study, we introduce a novel approach to achieve the merging of BICs at the edges of the fourth stop band, which opens at the third-order $\Gamma$ point, in one-dimensional leaky-mode photonic lattices. Photonic band gaps and BICs arise from periodic modulations in lattice parameters. However, near the third-order $\Gamma$ point, out-of-plane radiation arises by the first and second Fourier harmonic components in the lattice parameters. Accidental BICs can emerge at specific $k$ points where an optimal balance exists between these two Fourier harmonic components. We demonstrate that these accidental BICs are topologically stable, and their positions in momentum space can be precisely controlled by adjusting a specific lattice parameter that influences the strength of the first and second Fourier harmonic components. Furthermore, we show that accidental BICs can be merged at the third-order $\Gamma$ point, with or without a symmetry-protected BIC, by finely adjusting this specific lattice parameter while keeping other parameters constant.
\end{abstract}
\maketitle

\section{Introduction}

The capability to achieve high-$Q$ resonances in limited regions is highly beneficial for a wide range of photonic applications. Optical bound states in the continuum (BICs) are unique eigensolutions of electromagnetic wave equations in non-Hermitian photonic structures \cite{Marinica2008,Hsu2016,Koshelev2019}. BICs are exceptional because they can localize light within photonic structures with theoretically infinite lifetimes, even though their energy levels fall within a range that allows radiation to escape. Owing to their extraordinary ability to generate ultrahigh-$Q$ resonances, which significantly enhance light-matter interactions, BICs have been extensively investigated in recent years \cite{Kodigala2017,Bulgakov2017,YXGao2017,Minkov2018,ZLiu2019,Koshelev2020,SGLee2021-3,SMurai2022,MKang2023,HZhong2023,SHan2023,LLiu2023,SGLee2024-1}. BICs have been found in diverse photonic systems, including metasurfaces \cite{Koshelev2018,Kupriianov2019,Gorkunov2020,TCTan2021}, photonic crystals \cite{CWHsu2013,YYang2014,JWang2020}, fiber Bragg gratings \cite{XGao2019}, and plasmonic structures \cite{Azzam2018,BinAlam2021,YZhou2022}. BICs can be created through various physical mechanisms, such as parameter tuning \cite{Friedrich1985,Bulgakov2008,Rybin2017,LYuan2020}, symmetry incompatibility \cite{SGLee2019-1,Plotnik2011,FZhang2023,LHuang2023}, topological charge evolution \cite{WYE2020,TYoda2020,XYin2020,YZeng2021,XYin2023}, environmental engineering \cite{Cerjan2019}, parity-time symmetry \cite{QJSong2020,QSong2023}, and inverse design \cite{Gladyshev2023,WHuang2024}. Theoretically, BICs can confine resonances in photonic structures with infinite radiative $Q$ factors. However, in practical experiments, the extent of light confinement is limited by unwanted scattering losses, often caused by fabrication imperfections. Consequently, the observed resonances exhibit merely finite $Q$ factors.

Periodic planar photonic structures, such as one-dimensional (1D) gratings and two-dimensional (2D) photonic crystal slabs, can exhibit various types of BICs, if the lattice structures adhere to time-reversal symmetry, up-down mirror symmetry, and appropriate rotational symmetry \cite{LNi2016,Bulgakov2018,SGLee2020-2,SGLee2020-1,MSHwang2022,Neale2021}. Owing to their topological nature, symmetry-protected BICs, accidental BICs, and Friedrich-Wintgen BICs exhibit robustness against variations in lattice parameters, as long as the system’s symmetries remain intact \cite{BZhen2014}. Recent studies have demonstrated that in both 1D and 2D photonic lattice slabs, it is feasible to merge multiple BICs either at or away from the lattice’s second-order $\Gamma$ point \cite{JJin2019,MKang2022-2,MKang2022-1,SYu2024,SGLee2023}. At the $\Gamma$ point, accidental BICs can merge with symmetry-protected BICs \cite{SGLee2024-2}, whereas at off-$\Gamma$ points, they can merge with Friedrich-Wintgen BICs \cite{MKang2022-2}. This merging of BICs is crucial for practical applications, as it offers a promising strategy to minimize undesired out-of-plane scattering losses caused by fabrication imperfections. 

\begin{figure*}[t]
\centering\includegraphics[width=17.5 cm]{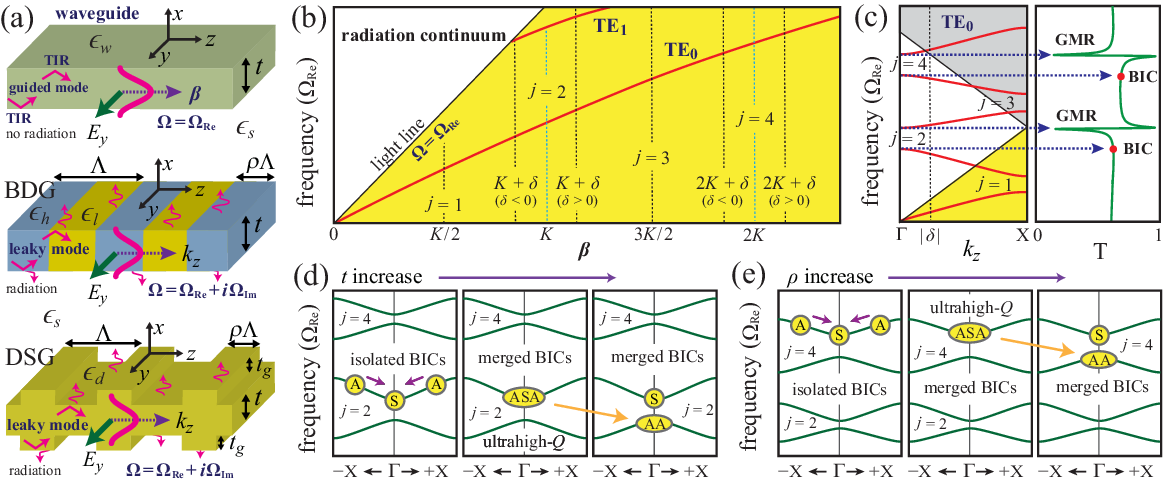}
\caption {\label{fig1} (a) Schematics of the homogeneous dielectric waveguide and two representative 1D photonic lattices used to study the merging of TE-polarized BICs at the third-order $\Gamma$ point. (b) Dispersion curves for $\mathrm{TE}_{0}$ and $\mathrm{TE}_{1}$ modes in a homogeneous dielectric waveguide. (c) Conceptual illustrations of photonic band structures and the transmission curve through 1D photonic lattices. Comparison of the merging of BICs at the second-order (d) and third-order (e) $\Gamma$ points. The letters S and A denote symmetry-protected BICs and accidental BICs, respectively.}
\end{figure*}

To date, the merging of BICs has typically required variations in slab thickness. Moreover, previously identified merging of BICs has generally occurred at or near the second-order $\Gamma$ point, associated with the second stop band. In principle, not only the second stop bands but also the fourth stop bands, which open at the third-order $\Gamma$ point, are capable of exhibiting BICs within the subwavelength domain \cite{SGLee2021-2}. However, studies on BICs in the fourth stop bands are scant. In this paper, we present a novel mechanism to merge multiple BICs at the third-order $\Gamma$ point. Photonic lattices exhibit fascinating topological physical phenomena, including BICs, exceptional points, and unidirectional light emission, all of which are attributed to the periodic modulation in lattice parameters. We show that near the third-order $\Gamma$ point, out-of-plane radiation arises from the interplay between the first and second Fourier harmonic components. This interaction gives rise to the emergence of accidental BICs at specific $k$ points, where there is an optimal balance between the first and second Fourier harmonic components. The position of these accidental BICs in momentum space can be precisely controlled by adjusting a specific lattice parameter that affects the strength of the first and second Fourier harmonic components while keeping other lattice parameters constant. Furthermore, we demonstrate that accidental BICs can be merged at the third-order $\Gamma$ point, with or without the presence of a symmetry-protected BIC, by carefully fine-tuning the strength of the first and/or second Fourier harmonic comonents.

\section{Lattice structure and perspective}

Figure~\ref{fig1}(a) depicts two representative 1D photonic lattices used to study the merging of BICs at the third-order $\Gamma$ point in this study. For comparison, a homogeneous slab waveguide composed of a core material with a dielectric constant $\epsilon_w$ is also presented. The homogeneous waveguide and the two photonic lattices are surrounded by a medium characterized by a dielectric constant $\epsilon_s$. The binary dielectric grating (BDG) consists of alternating materials with high ($\epsilon_h$) and low ($\epsilon_l$) dielectric constants. It features a lattice constant denoted by $\Lambda$, a grating layer thickness of $t$, and widths of the high dielectric constant sections expressed as $\rho\Lambda$. The average dielectric constant, denoted by $\epsilon_0 = \rho \epsilon_h + (1 - \rho) \epsilon_l$, is greater than $\epsilon_s$. The double-sided grating (DSG) features a slab waveguide layer with grating layers attached to both interfaces. The waveguide layer and attached grating layers are made of a dielectric material with a dielectric constant $\epsilon_d > \epsilon_s$. The thickness of the slab waveguide layer is $t$, and the thickness and width of the attached grating layers are $t_g$ and $\rho\Lambda$, respectively.

\begin{figure*}[]
\centering\includegraphics[width=17.5 cm]{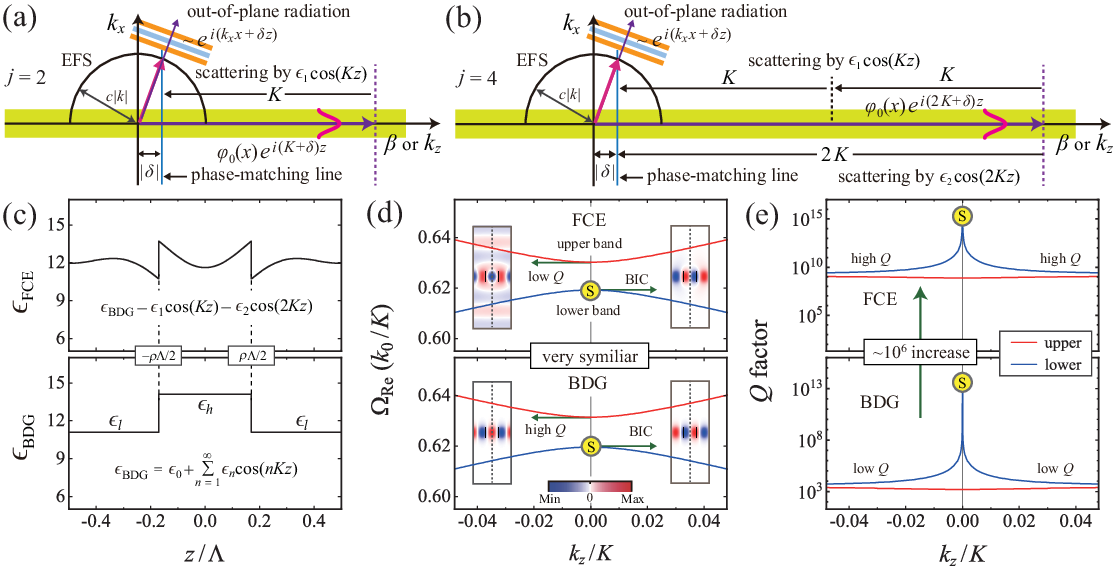}
\caption {\label{fig2} (a) Conceptual illustration of the out-of-plane radiation in the vicinities of the second ($j=2$) and fourth ($j=4$) stop bands. (b) Comparison between the key properties of the FCE metasurface and the conventional BDG. (c) Dielectric functions with respect to $z$. (d) FEM-simulated dispersion curves near the third-order $\Gamma$ point. Insets with blue and red colors illustrate the spatial electric field ($E_{y}$) distributions of band edge modes on the $y=0$ plane. (e) Simulated radiative $Q$ factors of the upper and lower dispersion curves. The structural parameters used in the FEM simulations were $t=0.42~\Lambda$, $\epsilon_{0}=3.48^2$, $\epsilon_{s}=1.46^2$, $\Delta \epsilon = 3$, and $\rho = 0.34$.}
\end{figure*}

In the homogeneous waveguide, characterized by $\epsilon_w > \epsilon_s$, total internal reflection (TIR) occurs at the interfaces between the core ($\epsilon_w$) and cladding ($\epsilon_s$) media. This TIR results in the waveguide modes exhibiting purely real eigenfrequencies ($\Omega = \Omega_{\mathrm{Re}}$) and propagating along the waveguide without any out-of-plane radiation loss \cite{Agrawal2004}. As illustrated in Fig.~\ref{fig1}(b), the dispersion curves for these transverse electric (TE) polarized waveguide modes are situated within the yellow region beneath the light line. The spatial electric field distribution of the $\mathrm{TE}_0$ mode satisfies $E_y(x,z) = \varphi_0(x) \exp{(i\beta z)}$, where $\varphi_0(x)$ represents the transverse profile of the $\mathrm{TE}_0$ mode. The representative 1D lattices also support TE polarized guided modes as $\epsilon_0$ and $\epsilon_d$ are greater than $\epsilon_s$. However, owing to the periodic modulations of the lattice parameters, all Bloch modes in the photonic lattice are represented within the irreducible Brillouin zone, as depicted in Fig.~\ref{fig1}(c). Photonic band gaps arise at the Bragg condition defined by $k_z = jK/2$, where $K = 2\pi/\Lambda$ is the magnitude of the grating vector and $j$ is an integer. In the vicinities of the second ($j=2$) and fourth ($j=4$) stop bands, located in the white region, guided modes exhibit complex eigenfrequencies $\Omega = \Omega_{\mathrm{Re}} + i\Omega_{\mathrm{Im}}$ and facilitate interesting leaky-wave phenomena such as BICs and guided-mode resonances (GMRs) through zero-order diffraction \cite{YDing2007}. Conversely, the first stop band ($j=1$) in the yellow region does not involve leaky-wave effects because guided modes are protected by TIR. Additionally, the third stop band ($j=3$) in the gray region is less practical, as it produces unwanted higher-order diffracted waves along with the zero-order waves.

Recent studies have extensively investigated the merging of BICs at the second-order $\Gamma$ point. As conceptually illustrated in Fig.~\ref{fig1}(d), accidental BICs can merge with symmetry-protected BICs at the edges of the second stop bands through variations in the thickness of photonic lattices. Although the fourth stop bands also have the potential to demonstrate BICs within the subwavelength domain, to the best of our knowledge, the merging of BICs at the third-order $\Gamma$ point has not yet been reported. In this study, we demonstrate that accidental BICs can merge at the third-order $\Gamma$ point by adjusting the lattice parameter $\rho$ while keeping other lattice parameters constant, as illustrated in Fig.~\ref{fig1}(e). The 1D photonic lattices depicted in Fig.~\ref{fig1}(a) are generally capable of supporting multiple TE-polarized guided modes, with each of these $\mathrm{TE}_{n \geq 0}$ modes potentially exhibiting multiple band gaps \cite{Magnusson2009}. For simplicity and clarity, our analysis specifically focuses on the merging of BICs at the edges of the fourth stop band, induced by the fundamental $\mathrm{TE}_0$ mode.


At the interfaces of 1D photonic lattices, as illustrated in Figs.~\ref{fig2}(a) and \ref{fig2}(b), out-of-plane radiation can be visualized by examining the continuity of the tangential components of wavevectors across the interfaces with the equifrequency surface (EFS) \cite{Notomi2000}. For the representative 1D photonic lattices with in-plane $C_2$ symmetry shown in Fig.~\ref{fig1}(a), the spatially periodic dielectric function can be represented by the Fourier series \cite{Inoue2004}:
\begin{equation}\label{epsilon}
\epsilon(x,z) = \epsilon_{0}(x) + \sum_{n=1}^{\infty} \epsilon_{n}(x) \cos (nKz).
\end{equation}
In this expression, $\epsilon_{n \geq 1}(x) = (2\Delta \epsilon / n\pi) \sin (n\pi \rho)$ when $x \in [-t, 0]$, while $\epsilon_{n}(x) = 0$ when $x \notin [-t, 0]$. A propagating wave, represented by $E_{y}(x,z) = \varphi_0(x) \exp{(i\beta z)}$, interacts, in principle, with an infinite number of Fourier harmonic components $\epsilon_{n} \cos (nKz)$, each with a period of $\Lambda/n$. As conceptually illustrated in Fig.~\ref{fig2}(a), however, near the second-order $\Gamma$ point, out-of-plane radiation is solely induced by the first Fourier harmonic component, $\epsilon_{1} \cos (Kz)$. Guided modes can fulfill the phase-matching condition due to the additional momentum $K$ imparted by the first  Fourier harmonic component. Properties of leaky guided modes near the second stop bands have been extensively explored in prior research \cite{SGLee2019-2}. Moreover, it has been demonstrated that Fourier-component-engineered (FCE) metasurfaces, that do not possess the first Fourier harmonic component, can support the continuous high-$Q$ bound states near the second stop bands \cite{SGLee2021-1}. Accidental BICs emerge at generic $\Gamma$ $k$ points when out-of-plane radiation, driven by the first Fourier harmonic component, vanishes. At the second-order $\Gamma$ point, the merging of multiple BICs is feasible since accidental BICs are topologically protected and shift along dispersion curves as the thickness of the lattice changes \cite{BZhen2014}. Prior studies have highlighted that varying the thickness is crucial for achieving the merging of BICs at the second-order $\Gamma$ point \cite{SGLee2023,SGLee2024-2}. In contrast, as illustrated in Fig.~\ref{fig2}(b), near the third-order $\Gamma$ point, both the first and second Fourier harmonic components, $\epsilon_{1} \cos (Kz)$ and $\epsilon_{2} \cos (2Kz)$, contribute to out-of-plane radiation simultaneously. Phase matching condition can be satisfied due to the additional momentum $2K$ provided by both $\epsilon_{1} \cos (Kz)$ and $\epsilon_{2} \cos (2Kz)$. 

The effect of the first and second Fourier harmonic components on out-of-plane radiation can be verified by comparing the dispersion properties of the conventional BDG with those of corresponding FCE metasurface, that do not have both the first and second Fourier harmonic components. As shown in Fig.~\ref{fig2}(c), the conventional BDG has a step-like dielectric function given by $\epsilon_{\rm{BDG}} = \sum_{n=0}^{\infty} \epsilon_{n} \cos (nKz)$, whereas the FCE metasurface features a more complex dielectric function, $\epsilon_{\rm{FCE}} = \epsilon_{\rm{BDG}} - \epsilon_{1} \cos (Kz) - \epsilon_{2} \cos (2Kz)$. As illustrated in Fig.~\ref{fig2}(d), near the third-order $\Gamma$ point, the dispersion curves for both the BDG and the FCE metasurface display similarities, and the spatial electric field ($E_y$) distributions, depicted in the insets, indicate that symmetry-protected BICs occur at the lower band edges of both structures. However, the radiative $Q$ factors as illustrated in Fig.~\ref{fig2}(e), highlight a notable distinction. Within the computational range of $|k_z|\leq 0.045~K$, the $Q$ factors of Bloch modes in the FCE metasurface are approximately $10^6$ times greater than those in the conventional BDG, despite identical lattice parameters apart from the dielectric function profile. This noticeable enhancement in the $Q$ factors demonstrates that near the fourth stop bands, out-of-plane leakage loss is primarily originate from the first and second Fourier harmonic components.

From Fig.~\ref{fig2}, in this study, we propose that near the third-order $\Gamma$ point, the total radiating wave from conventional BDGs results from the superposition of two radiating waves, each originating from the first and second Fourier harmonic components, respectively. While the second Fourier coefficient $\epsilon_{2} = (\Delta \epsilon / \pi) \sin (2\pi \rho)$ changes sign from $+$ to $-$ as $\rho$ varies, the first Fourier coefficient $\epsilon_{1} = (2\Delta \epsilon / \pi) \sin (\pi \rho)$ consistently remains positive. Previous studies have analytically and numerically demonstrated that the first and second Fourier harmonic components can exhibit constructive or destructive interactions depending on the value of $\rho$ \cite{SGLee2019-1,SGLee2019-2}. We conjecture here that in the vicinity of the third-order $\Gamma$ point, radiating waves due to $\epsilon_{1} \cos (Kz)$ and $\epsilon_{2} \cos (2Kz)$ can destructively interfere depending on the lattice parameter $\rho$. With a specific value of $\rho$, accidental BICs can emerge at particular $k$ points where the first and second Fourier harmonic components are optimally balanced, leading to perfect destructive interference of the two radiating waves. Due to their topological characteristics, these accidental BICs can move along the dispersion curves and merge at the third-order $\Gamma$ point with continuous variation of $\rho$, while maintaining other lattice parameters constant.

\begin{figure*}[]
\centering\includegraphics[width=17.5 cm]{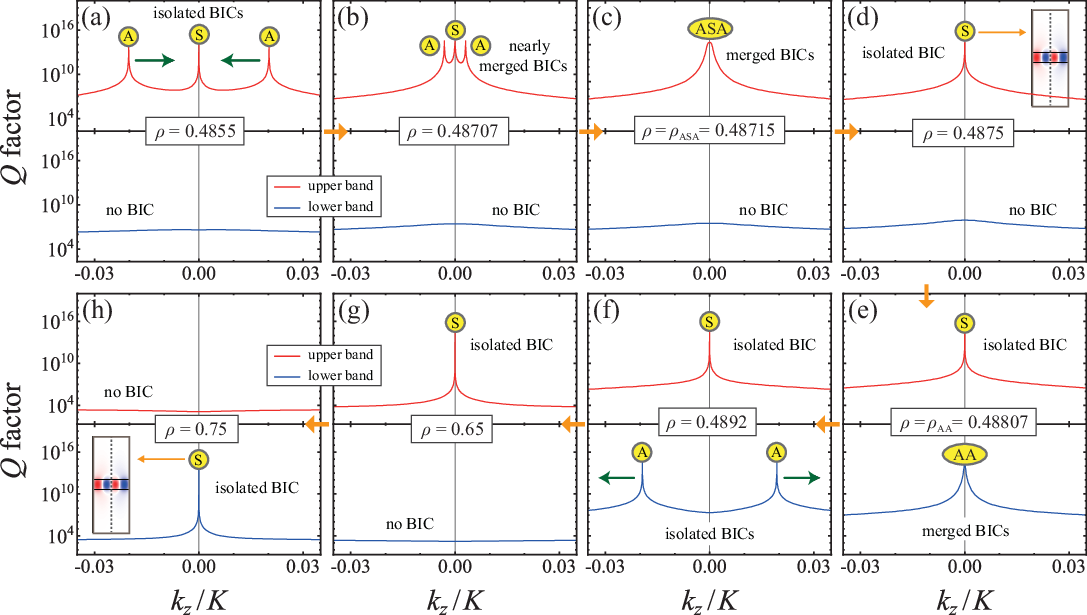}
\caption {\label{fig3} Calculated radiative $Q$ factor curves near the third-order $\Gamma$ point for varying lattice parameter $\rho$: (a) $0.4855~\Lambda$, (b) $0.48707~\Lambda$, (c) $0.48715~\Lambda$, (d) $0.4875~\Lambda$, (e) $0.48807~\Lambda$, (f) $0.4892~\Lambda$, (g) $0.65~\Lambda$, and (h) $0.75~\Lambda$. As the value of $\rho$ increases, accidental BICs shift along the dispersion curves, leading to the merging of BICs at the upper band edge. The merged state of BICs transitions from the upper to the lower dispersion curve as $\rho$ increases. For the FEM simulations, the structural parameters used were $t=0.42~\Lambda$, $\epsilon_{0}=3.48^2$, $\epsilon_{s}=1.46^2$, and $\Delta \epsilon=3$.}
\end{figure*}

\section{Results and discussion}
\subsection{Merging BICs in BDGs}

Figure~\ref{fig2}(e) shows that accidental BICs do not appear near the third-order $\Gamma$ point within the computational range of $|k_z|\leq 0.045~K$ when $\rho = 0.34$. We now demonstrate the merging of BICs in conventional BDGs by examining the evolution of the $Q$ factor curves under variation of $\rho$. Figure~\ref{fig3} presents the radiative $Q$ factor curves for both the upper and lower band branches as a function of $k_z$. When $\rho = 0.4855$, as depicted in Fig.~\ref{fig3}(a), the BDG exhibits two accidental BICs in the upper band at off-$\Gamma$ $k$ points, along with a symmetry-protected BIC at the third-order $\Gamma$ point. Due to the in-plane $C_2$ symmetry of the BDG, these two accidental BICs simultaneously appear at $k_z = \pm 0.02017~K$. As the value of $\rho$ increases beyond $0.4855$, these accidental BICs progressively shift downward along the upper dispersion curve, moving closer to the third-order $\Gamma$ point. At $\rho = 0.48707$, as depicted in Fig.~\ref{fig3}(b), a nearly merged state of two accidental BICs and a symmetry-protected BIC is observed in the upper band. Upon further increasing $\rho$ to $\rho_{\rm{ASA}}=0.48715$, as shown in Fig.~\ref{fig3}(c), the accidental BICs merge with the symmetry-protected BIC at the third-order $\Gamma$ point, forming a single, pronounced peak in the $Q$ factor curve. Compared to the isolated symmetry-protected BICs or accidental BICs, the merged state at $\rho=\rho_{\rm{ASA}}$ exhibits significantly improved radiative $Q$ factors over a wide range of wavevectors.

As the value of $\rho$ further increases beyond $\rho_{\rm{ASA}}=0.48715$, the merged state of BICs disappears, leaving only a single symmetry-protected BIC at the upper band edge. The inset in Fig.~\ref{fig3}(d) displays the asymmetric $E_y$ field distribution characteristic of the symmetry-protected BIC. Furthermore, Fig.~\ref{fig3}(d) indicates that at $\rho = 0.4875$, no accidental BICs are observed in either the upper or lower bands. However, as the value of $\rho$ continues to increase to $\rho_{\rm{AA}}=0.48807$, simultaneous peaks appear in the $Q$ factor curves for both the upper and lower band edges, as shown in Fig.~\ref{fig3}(e). The peak at the upper band edge is due to the presence of a symmetry-protected BIC, while the prominent peak at the lower band edge results from the merged state of accidental BICs. Previous research has shown that the merged state of accidental BICs transitions from the upper to the lower band edge of the second stop band as the thickness $t$ increases \cite{SGLee2023}. Similarly, in this study, we observe that the interband transition of the merged state of accidental BICs across the fourth stop band is facilitated by varying the lattice parameter $\rho$ while keeping other parameters constant. Upon further increasing $\rho$ beyond $\rho_{\rm{AA}}=0.48807$, as illustrated in Fig.~\ref{fig3}(f) with $\rho = 0.4892$, the merged state at the lower band edge splits into two distinct peaks, corresponding to accidental BICs. When $\rho = 0.65$, Fig.~\ref{fig3}(g) shows that the two accidental BICs disappear from the lower band, while a symmetry-protected BIC remains at the upper band edge. Conversely, when $\rho = 0.75$, Fig.~\ref{fig3}(h) reveals that only a symmetry-protected BIC appears at the lower band edge, with no accidental BICs present in either the upper or lower bands.

\begin{figure}[b]
\centering\includegraphics[width=7.5 cm]{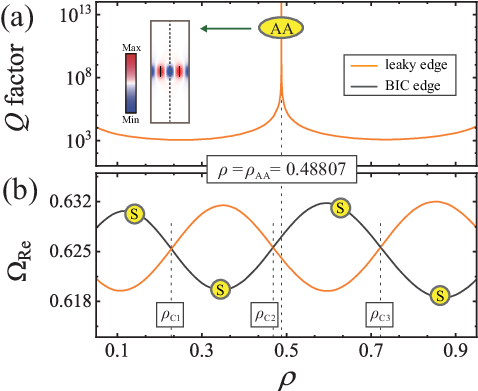}
\caption {\label{fig4} (a) Calculated radiative $Q$ factors of the band edge modes, which have symmetric electric field distributions. The inset shows that the merged state with two accidental BICs does not radiate outside the lattice. The radiative $Q$ factors for BICs are not shown as their values exceed $10^{14}$. (b) Band edge frequencies at the fourth stop band as a function of $\rho$. }
\end{figure}

\begin{figure*}[t]
\centering\includegraphics[width=17.5 cm]{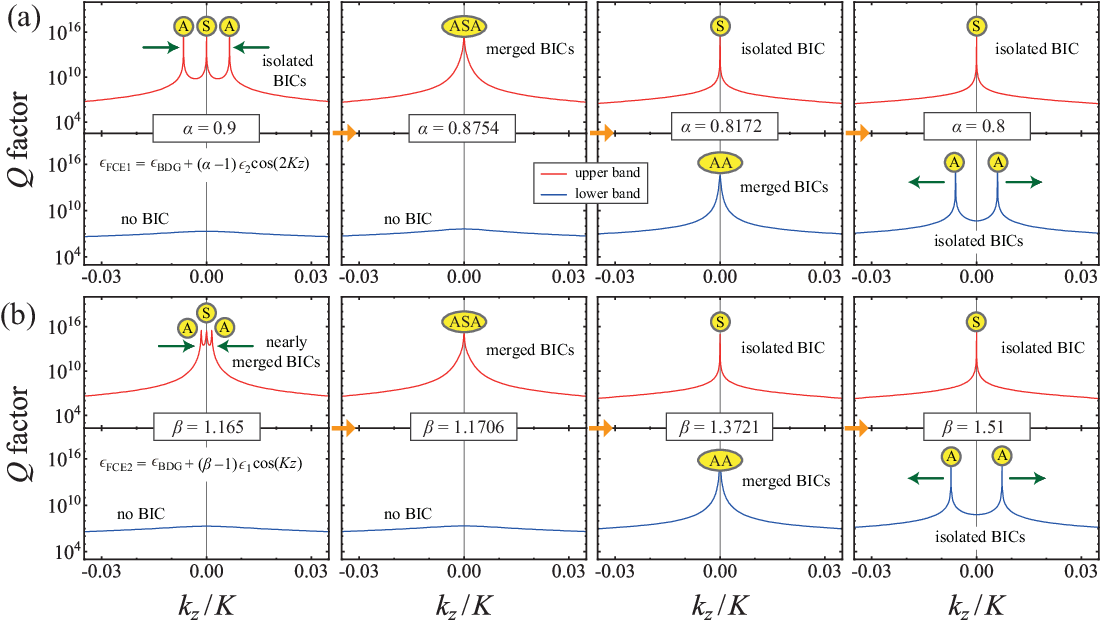}
\caption {\label{fig5} Calculated radiative $Q$ factor curves near the third-order $\Gamma$ point for FCE metasurfaces with spatial dielectric functions given by (a) $\epsilon_{\rm{FCE1}} = \epsilon_{\rm{BDG}} + (\alpha - 1) \epsilon_{2} \cos (2Kz)$ and (b) $\epsilon_{\rm{FCE2}} = \epsilon_{\rm{BDG}} + (\beta -1)\epsilon_{1} \cos (Kz)$. By controlling the coefficients $\alpha$ and $\beta$ while keeping other lattice parameters constant, the strength of the first and second Fourier harmonic components, respectively, can be precisely adjusted.}
\end{figure*}

Figure~\ref{fig3} shows that two accidental BICs can be merged at the third-order $\Gamma$ point with a symmetry-protected BIC when $\rho = \rho_{\rm{ASA}}$, and without a symmetry-protected BIC when $\rho = \rho_{\rm{AA}}$. The merged states of two accidental BICs at $\rho = \rho_{\rm{AA}}$ can be straightforwardly identified by calculating the radiative $Q$ factors of band edge modes possessing symmetric $E_y$ field distributions as a function of $\rho$. Figure~\ref{fig4}(a) demonstrates that at the third-order $\Gamma$ point, as $\rho$ varies from 0 to 1, the merging of accidental BICs occurs once at $\rho = \rho_{\rm{AA}}=0.48807$. Generally, band edge modes with symmetric $E_y$ field distributions radiate outside photonic lattices. However, the inset in Fig.~\ref{fig4}(a) illustrates that the merged state with accidental BICs does not radiate outside the lattice, despite having symmetric field distributions. The merged state with a symmetry-protected BIC at $\rho_{\rm{ASA}}$, a nearly merged state, and isolated accidental BICs near the third-order $\Gamma$ point can be easily created by using $\rho_{\rm{AA}}$ as a reference. We also plotted the eigenfrequencies of the two band edge modes in Fig.~\ref{fig4}(b). As $\rho$ varies, the fourth stop band exhibits closed band states three times at $\rho = \rho_{\rm{C1}}=0.22756$, $\rho_{\rm{C2}}=0.46843$, and $\rho_{\rm{C3}}=0.72239$. Previous studies have shown that the size of the fourth stop band is primarily dependent on the fourth Fourier harmonic component, $\epsilon_{4} \cos (4Kz)$, where $\epsilon_{4} = (\Delta \epsilon / 2\pi) \sin (4\pi \rho)$ \cite{SGLee2021-2,Yariv1984}. Other Fourier harmonic components may contribute to the band gap size auxiliary. Consequently, the fourth stop band closes near three points: $\rho = 0.25$, $\rho = 0.5$, and $\rho = 0.75$, where $\epsilon_{4}$ becomes zero. Before and after the band gap closure, the relative positions of symmetry-protected BICs and leaky modes are reversed. Hence, a symmetry-protected BIC can be found at either the upper or lower band edge depending on the value of $\rho$. 

\begin{figure*}[t]
\centering\includegraphics[width=17.5 cm]{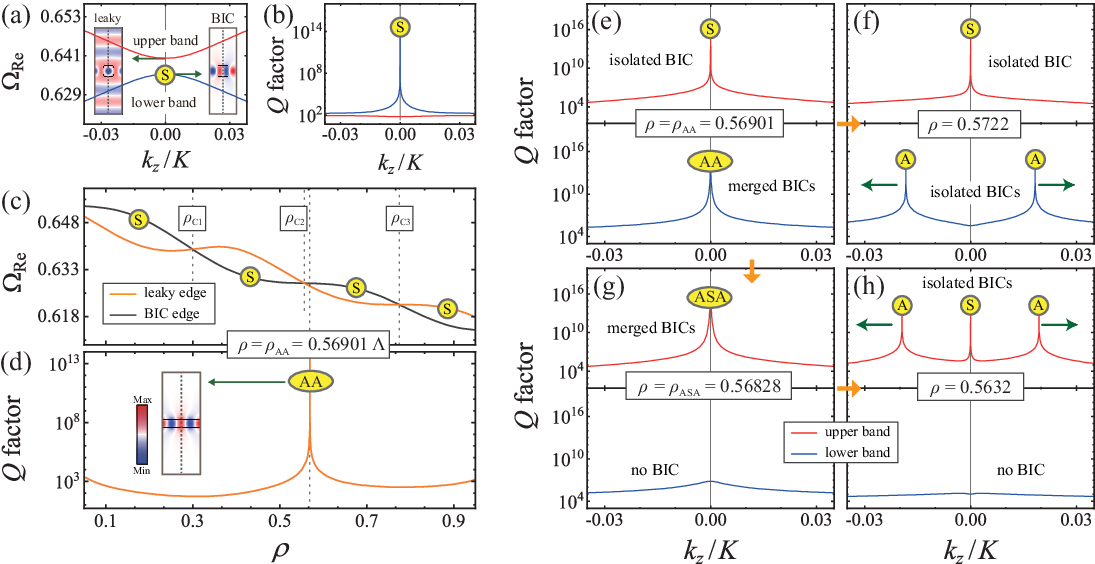}
\caption {\label{fig6} FEM-simulated dispersion curves (a) and radiative $Q$ factors (b) near the third-order $\Gamma$ point. In the simulations, we used lattice parameters $\rho = 0.35$, $t = 0.3~\Lambda$, $t_g = 0.1~\Lambda$, $\epsilon_d = 3.48^2$, and $\epsilon_s = 1$. Calculated radiative $Q$ factor curves for both the upper and lower bands when (e) $\rho = 0.56901$, (f) $\rho = 0.5722$, (g) $\rho = 0.56828$, and (h) $\rho = 0.5632$. Two accidental BICs are merged with and without a symmetry-protected BIC at $\rho = \rho_{\rm{ASA}} = 0.5722$ and $\rho = \rho_{\rm{AA}} = 0.56901$, respectively.}
\end{figure*}

In this study, we propose that near the third-order $\Gamma$ point, accidental BICs emerge due to the optimal balance between the first and second Fourier harmonic components. As shown in Fig.~\ref{fig3}(a), two accidental BICs appear at $k_z = \pm 0.02017~K$ when $\rho = 0.4855$. As the value of $\rho$ (less than 0.5) progressively increases from $0.4855$, the strength of the first Fourier harmonic component increases, while that of the second Fourier harmonic component decreases. Accidental BICs can move along the dispersion curves because the BIC points, where the first and second Fourier harmonic components are optimally balanced, shift with the increase in $\rho$. To demonstrate our conjecture, we consider a conceptual FCE metasurface with a spatial dielectric function given by $\epsilon_{\rm{FCE1}} = \epsilon_{\rm{BDG}} - \epsilon_{2} \cos (2Kz) + \alpha \epsilon_{2} \cos (2Kz)$, where the coefficient $\alpha$ is introduced to control the strength of the second Fourier harmonic component. The proposed FCE metasurface has the second Fourier harmonic component $\alpha \epsilon_{2} \cos (2Kz)$, while the conventional BDG has $\epsilon_{2} \cos (2Kz)$. All other Fourier harmonic components for the BDG and the FCE metasurface are the same. As the value of $\alpha$ gradually decreases from 1, while $\rho$ is fixed at $0.4855$, it is reasonably expected that the two accidental BICs will shift downward along the upper dispersion curve, moving closer to the third-order $\Gamma$ point. Figure~\ref{fig5}(a) illustrates the evolution of BICs with varying coefficients of $\alpha$. When $\alpha = 0.9$, compared to Fig.~\ref{fig3}(a) where $\alpha = 1$, the two accidental BICs move closer to the symmetry-protected BIC at the third-order $\Gamma$ point. At $\alpha = 0.8754$, a merged state comprising two accidental BICs and a symmetry-protected BIC is observed at the upper band edge. At both $\alpha = 0.9$ and $\alpha = 0.8754$, no BIC is present in the lower band. As $\alpha$ further decreases to $0.8172$, a symmetry-protected BIC and the merged state of two accidental BICs are simultaneously observed at the upper and lower band edges, respectively. Upon further decreasing $\alpha$ to 0.8, the merged state at the lower band edge splits into two isolated peaks corresponding to accidental BICs.

We consider another FCE metasurface with a spatial dielectric function given by $\epsilon_{\rm{FCE2}} = \epsilon_{\rm{BDG}} - \epsilon_{1} \cos (Kz) + \beta \epsilon_{1} \cos (Kz)$. This FCE metasurface has the first Fourier harmonic component $\beta \epsilon_{1} \cos (Kz)$, whereas the conventional BDG has $\epsilon_{1} \cos (Kz)$. As the value of $\beta$ gradually increases from 1, while $\rho$ is fixed at $0.4855$, the strength of the first Fourier harmonic component increases. Consequently, the position of accidental BICs can be adjusted by tuning the value of $\beta$. Figure~\ref{fig5}(b) illustrates the evolution of BICs with varying $\beta$. When $\beta = 1.165$ and $\beta = 1.1706$, nearly merged BICs and a merged state of BICs are observed in the upper band. As $\beta$ further increases to $1.3721$, a symmetry-protected BIC and the merged state of two accidental BICs are simultaneously observed at the upper and lower band edges, respectively. Upon further increasing $\beta$ to $1.51$, the merged state at the lower band edge splits into two isolated peaks corresponding to accidental BICs. Figure~\ref{fig5} demonstrates that by controlling the strength of the first and/or second Fourier harmonic components, while keeping the strength of other Fourier harmonic components unchanged, accidental BICs can be repositioned in the momentum space and merged at the third-order $\Gamma$ point.

\subsection{Merging BICs in DSGs}

We next investigate the merging of BICs at the third-order $\Gamma$ point in DSGs. As illustrated in Fig.~\ref{fig1}(a), DSGs consist of a core waveguide slab and two BDGs. The spatial dielectric constant modulations in the BDGs enable DSGs to exhibit photonic band gaps and BICs. Therefore, it is reasonable to anticipate that merging BICs can be achieved by adjusting the width of the high dielectric constant regions, $\rho \Lambda$. Figures~\ref{fig6}(a) and \ref{fig6}(b) present simulated dispersion curves and radiative $Q$ factors, respectively, for the DGS with lattice parameters $\rho = 0.35$, $t = 0.3~\Lambda$, $t_g = 0.1~\Lambda$, $\epsilon_d = 3.48^2$, and $\epsilon_s = 1$. As depicted in Fig.~\ref{fig6}(a), the fourth band gap opens at the third-order $\Gamma$ point. The spatial electric field distributions in the insets of Fig.~\ref{fig6}(a) reveal that one of the band edge modes, characterized by an asymmetric field distribution, forms a nonleaky symmetry-protected BIC, whereas the other mode, with a symmetric field distribution, radiates out of the lattice. The radiative $Q$ factors depicted in Fig.~\ref{fig6}(b) show that for $\rho = 0.35$, only a symmetry-protected BIC is present at the lower band edge within the computational range of $|k_z| \leq 0.04~K$. No accidental BIC is observed in either the upper or lower bands. Figure~\ref{fig6}(c) depicts the simulated eigenfrequencies of band edge modes at the third-order $\Gamma$ point in DSGs. As $\rho$ varies from 0 to 1, the fourth band gap closes three times at $\rho = \rho_{\rm{C1}}=0.29961$, $\rho_{\rm{C2}}=0.55626$, and $\rho_{\rm{C3}}=0.77477$. Before and after the band gap closure, the relative positions of symmetry-protected BICs and leaky modes are reversed. In BDGs, as illustrated in Fig.~\ref{fig5}(a), as $\rho$ varies from 0 to 1, two accidental BICs merge once at $\rho = \rho_{\rm{AA}}=0.48807$ without the presence of a symmetry-protected BIC. Similarly, in DSGs, Fig.~\ref{fig6}(c), which depicts the radiative $Q$ factors of band edge modes with symmetric electric field distributions, reveals that two accidental BICs merge only once at $\rho = \rho_{\rm{AA}} = 0.56901$.

We also investigated the radiative $Q$ factors for the upper and lower bands as a function of $k_z$ near the fourth stop band under varying $\rho$. The four selected results are presented in Figs.~\ref{fig6}(e)--\ref{fig6}(h). When $\rho = \rho_{\rm{AA}} = 0.56901$, as shown in Fig.~\ref{fig6}(e), a symmetry-protected BIC is observed at the upper band edge, and the merged state of two accidental BICs is simultaneously observed at the lower band edge. As $\rho$ increases from $\rho_{\rm{AA}} = 0.56901$ to 0.5722, Fig.~\ref{fig6}(f) reveals that the merged state at the lower band edge splits into two isolated peaks corresponding to accidental BICs, while the symmetry-protected BIC remains at the upper band edge. Conversely, when $\rho$ decreases from $\rho_{\rm{AA}}$ to $\rho_{\rm{ASA}} = 0.56828$, Fig.~\ref{fig6}(g) demonstrates that a single merged state composed of two accidental BICs and a symmetry-protected BIC is observed at the upper band edge. As the two accidental BICs transition from the lower to the upper band as $\rho$ decreases, no BIC is observed in the lower band at $\rho = \rho_{\rm{ASA}}$. Upon further decreasing $\rho$ to 0.5632, Fig.~\ref{fig6}(h) shows that the merged state at the upper band edge splits into three isolated peaks corresponding to two accidental BICs and a symmetry-protected BIC.

\section{Conclusion}

In conclusion, we have conducted a comprehensive investigation into the merging of BICs at the third-order $\Gamma$ point in two representative 1D photonic lattices using rigorous FEM simulations. Our study reveals that near the fourth stop bands, which open at the third-order $\Gamma$ point, Bloch guided modes experience out-of-plane leakage loss induced by the first and second Fourier harmonic components in the lattice parameters. In 1D photonic lattices with in-plane $C_2$ symmetry and up-down mirror symmetry, a symmetry-protected BIC is observed at the edge of the fourth stop band. Additionally, accidental BICs can emerge at specific $k$ points, including the third-order $\Gamma$ point, where there is an optimal balance between the first and second Fourier harmonic components. In both BDGs and DSGs, the strength of these Fourier harmonic components is significantly influenced by the lattice parameter $\rho$, which represents the width of the high dielectric constant materials in gratings. With in-plane $C_2$ symmetry, two accidental BICs appear simultaneously in the momentum space. These accidental BICs are topologically stable and can traverse dispersion curves as $\rho$ varies. As $\rho$ gradually changes from zero to one, while keeping other lattice parameters constant, two accidental BICs can merge with a symmetry-protected BIC at $\rho = \rho_{\rm{ASA}}$ and without a symmetry-protected BIC at $\rho = \rho_{\rm{AA}}$. The merging of multiple BICs holds significant promise for practical applications, as it facilitates the creation of ultrahigh-$Q$ resonances, thereby substantially enhancing light-matter interactions. Our findings present a robust and effective method for manipulating electromagnetic waves by leveraging high $Q$ resonances, paving the way for advanced photonic devices and applications.

\acknowledgments
This research was supported by grants from the National Research Foundation of Korea, funded by the Ministry of Science and ICT (No. 2022R1A2C1011091) and the Ministry of Education (No. 2021R1I1A1A01060447). Additionally, this work was supported by a research grant from Kongju National University in 2021.

\end{document}